\documentclass[aps,prd,twocolumn,showpacs,showkeys]{revtex4-1}

\usepackage{euscript}
\usepackage{amssymb}
\usepackage{graphicx}
\usepackage{amsmath}
\usepackage{amsbsy}
\usepackage{amsthm}
\usepackage{bbm}
\usepackage{bm}
\usepackage{epsfig}
\usepackage{epstopdf}
\usepackage{dsfont}
\usepackage{color}
\usepackage[colorlinks]{hyperref}
\usepackage[figure,table]{hypcap}
\hypersetup{
	bookmarksnumbered,
	pdfstartview={FitH},
	citecolor={darkgreen},
	linkcolor={darkred},
	urlcolor={darkblue},
	pdfpagemode={UseOutlines}}
\definecolor{darkgreen}{RGB}{50,190,50}
\definecolor{darkblue}{RGB}{0,0,190}
\definecolor{darkred}{RGB}{238,0,0}
\usepackage{soul}

\newcommand{\tr}{\textnormal{Tr}}
\newcommand{\bra}[1]{\ensuremath{\left\langle\right. #1 \left.\right|}}
\newcommand{\ket}[1]{\ensuremath{\left|\right. #1 \left.\right\rangle}}
\newcommand{\fbra}[1]{\ensuremath{\left\langle\!\hspace*{-0.7pt}\left\langle\right.\right.\! #1 \!\left.\left.\right|\!\right|}}
\newcommand{\fket}[1]{\ensuremath{\left|\!\left|\right.\right.\! #1 \!\left.\left.\right\rangle\!\hspace*{-0.7pt}\right\rangle}}

\newcommand{\pr}{^{\prime}}
\newcommand{\prpr}{^{\prime\hspace*{-0.5pt}\prime}}

\DeclareMathOperator{\atanh}{atanh}

\usepackage{geometry}
\begin{document}

\title{Quantum gates and multipartite entanglement resonances realized by non-uniform cavity motion}

\date{October 2012}

\author{Nicolai Friis$^{1}$}
\email{pmxnf@nottingham.ac.uk}
\author{Marcus Huber$^{2}$}
\email{mazmh@bristol.ac.uk}
\author{Ivette Fuentes$^{1}$}
\email{ivette.fuentes@nottingham.ac.uk}
\thanks{Previously known as Fuentes-Guridi and Fuentes-Schuller.}
\author{David Edward Bruschi$^{1}$}
\email{pmxdeb@nottingham.ac.uk}
\affiliation{
$^{1}$School of Mathematical Sciences,
University of Nottingham,
University Park,
Nottingham NG7 2RD,
United Kingdom}
\affiliation{$^{2}$Department of Mathematics,
University of Bristol,
University Walk,
Clifton, Bristol BS8 1TW,
United Kingdom}

\begin{abstract}
We demonstrate the presence of genuine multipartite entanglement between
the modes of quantum fields in non-uniformly moving cavities. The
transformations generated by the cavity motion can be considered as
multipartite quantum gates.
We present two setups for which multi-mode entanglement can be generated
for bosons and fermions.
As a highlight we show that the genuine bosonic multipartite correlations
can be resonantly enhanced.
Our results provide fundamental insights into the structure of
Bogoliubov transformations and suggest strong links between quantum
information, quantum fields in curved spacetimes and gravitational
analogues by way of the equivalence principle.
\end{abstract}



\maketitle

\section{Introduction}\label{sec:intro}

During the past decade research in the area of relativistic quantum information has
addressed questions concerning the inherently relativistic aspects of quantum phenomena,
unveiling the close connections between effects in quantum field theories and quantum
information theory.

In this context several models to store and access bipartite quantum correlations in relativistic
settings have been proposed, including Unruh-DeWitt type detector models~\cite{HuLinLouko2012},
free field modes~\cite{BruschiLoukoMartin-MartinezDraganFuentes2010,DownesRalphWalk2012,
DraganDoukasMartin-MartinezBruschi2012} and cavity modes (see Refs.~\cite{DownesFuentesRalph2011,
BruschiFuentesLouko2012,FriisLeeBruschiLouko2012,FriisBruschiLoukoFuentes2012,FriisFuentes2012,
BruschiDraganLeeFuentesLouko2012} for details and Ref.~\cite{AlsingFuentes2012} for a recent review).

In these situations bipartite entanglement was extensively studied but, save for a few exceptions~\cite{AlsingFuentes-SchullerMannTessier2006,AdessoFuentes-SchullerEricsson2007,
AdessoFuentes-Schuller2009,HuberFriisGabrielSpenglerHiesmayr2011,SmithMann2012}, the
role of genuine multipartite correlations in these studies was marginal. However, multipartite
correlations are expected to feature in relativistic scenarios, e.g., in the transformation
of Minkowski modes to Rindler modes~\cite{BruschiLoukoMartin-MartinezDraganFuentes2010}. Moreover,
they become ever more prominent as the size and complexity of the systems in question increase.
In these scenarios the classification of genuine multipartite entanglement will allow for a more
detailed characterization of relativistic effects and may be used for high-sensitivity tests that
indicate the \emph{``quantumness''} of relativistic phenomena. In fact, the identification of
quantum correlations can be a keystone to the experimental verification of effects in quantum field
theory, such as the Hawking effect in analog fluid systems, see Ref.~\cite{RecatiPavloffCarusotto2009}.

Here we want to focus on the \emph{generation of genuine multipartite entanglement} by the
non-uniform motion of cavities that contain relativistic quantum fields. We employ the techniques
developed in Refs.~\cite{BruschiFuentesLouko2012,FriisLeeBruschiLouko2012}, for bosonic and fermionic
fields respectively, where the moving cavities follow worldtubes that are composed of segments of
inertial motion and uniform acceleration. In this framework we ask the questions: \emph{Can the
Bogoliubov transformations that are induced by the non-uniform motion generate genuine multipartite,
quantum correlations? If this is the case, can we quantify and/or classify the arising entanglement?}

We present two scenarios for which we can indeed create, and partially classify, such correlations,
thereby realizing \emph{quantum gates by motion}. First, in Sec.~\ref{sec:scenario A}, we consider
three individual cavities, labelled Alice, Rob and Charlie, that share pairwise bipartite entanglement
in their initially common rest-frame, before Rob's cavity undergoes non-uniform motion, see
Fig.~\ref{fig:multi cavity setup}. In the second scenario we consider Rob's cavity on its own in
Sec.~\ref{sec:scenario B}. In these scenarios the significantly different advantages of bosonic and
fermionic systems become apparent.

In the bosonic case the transformations induced by the cavity motion generate multi-mode entangling
quantum gates. Moreover, we show that the genuine multipartite character of the bosonic
entanglement can be enhanced resonantly by appropriate timing of the cavity's trajectory segments.
The qubit structure of the fermionic systems, on the other hand, allows for a clear classification
of the arising multipartite correlations. The cavity motion effectively acts as a multipartite
quantum gate producing Dicke states and $W$ states. We present explicit pure state decompositions for
these classifications in Secs.~\ref{subsec:scenario A fermions} and \ref{subsec:scenario B fermions},
respectively.

The calculation of the Bogoliubov transformations for the non-uniform cavity motion are performed
perturbatively in terms of a parameter $h$ that physically represents the product of the (rest-frame)
cavity width $\delta$ and the acceleration at the centre of the cavity. Using the Bogoliubov coefficients
obtained in this way we can perform all other computations analytically. For simplicity we restrict our
considerations to $(1+1)$-dimensional Minkowski spacetime where the metric tensor $\eta_{\mu\nu}$ has the
signature $(-,+)$. The setup can be naturally extended to accommodate additional spatial dimensions that
enter into the effective mass of the field via the corresponding transverse
momenta~\cite{BruschiFuentesLouko2012}. We use units where Planck's constant and the speed of light are
dimensionless, i.e., $\hbar=c=1$. By $O(x)$ we denote a quantity for which $O(x)/x$ remains bounded as $x$
goes to zero. We use an asterisk and a dagger to denote complex and Hermitian conjugation respectively.

\section{Multi-cavity entanglement}\label{sec:scenario A}

\begin{figure}
\centering
\vspace*{-4ex}%
\includegraphics[width=0.45\textwidth]{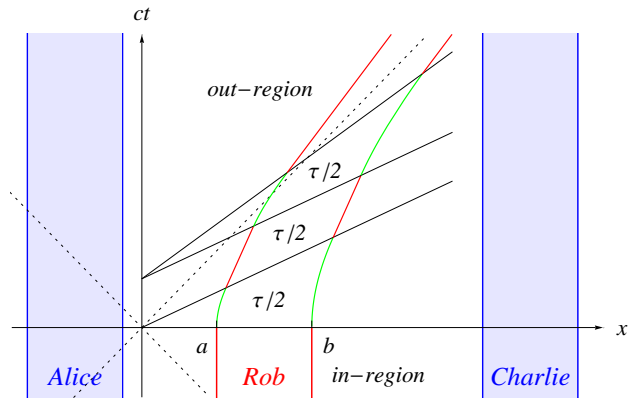}
\caption{
Space-time diagram of the multi-cavity setup: While Alice's and Charlie's cavities
remain at rest Rob's cavity undergoes non-uniform motion. Rob's cavity  of width
$\delta = (b - a)$ follows a worldtube that consists of alternating segments of
inertial motion (red) and uniform acceleration (green). The individual segments
in this example travel scenario are of equal duration $\tau/2$ and equal
acceleration $h/\delta$ as measured at the centre of Rob's cavity. The dashed lines
indicate the light cone at $(t,x)=(0,0)$.
\label{fig:multi cavity setup}}
\end{figure}

As the first scenario we consider three cavities, Alice, Rob and Charlie. Each cavity represents an
individual spacetime in which a relativistic quantum field is confined by appropriate boundary conditions.
We can assume without loss of generality that all three cavities are manufactured in the same way and that
they are initially (in the ``in-region'') at rest with respect to each other. At $t=0$ Rob's cavity starts
to accelerate linearly and uniformly. We assume that the worldtube of Rob's cavity consists of periods of
inertial motion and uniform acceleration such that the cavity remains rigid throughout the journey, i.e.,
the cavity length $\delta>0$, as measured by the co-moving observer, is constant. For the perturbative
treatment we further assume that the accelerations in all individual segments are small with respect to
the inverse cavity length $1/\delta$. Fig.~\ref{fig:multi cavity setup} shows the spacetime diagram of
this setup for a sample travel scenario.\\

\subsection{Scalar fields}\label{subsec:scenario A bosons}

Let us consider a real, scalar field $\phi$ of mass $m\geq0$ in Rob's initially inertial cavity in the
``in-region''. The field satisfies the Klein-Gordon equation $(-\Box+m^{2})\phi=0$, where $\Box$ is the
scalar D'Alambertian. We follow the procedure laid out in Ref.~\cite{BruschiFuentesLouko2012} and
represent the confinement to the cavity by imposing the Dirichlet boundary conditions
$\phi(t<0,a)=\phi(t<0,b)=0$. We thus obtain a discrete spectrum of mode functions $\phi_{n}$ in the
in-region, where the field can be decomposed as
\begin{align}
    \phi    &= \sum\limits_{n}\,\Bigl(\,\phi_{n}\,a_{n}\,+\,\phi_{n}^{*}\,a_{n}^{\dagger}\,\Bigr)\,.
    \label{eq:in region field}
\end{align}
The annihilation and creation operators, $a_{n}$ and $a_{n}^{\dagger}$, satisfy the usual commutation
relations $\left[a_{n},a_{m}^{\dagger}\right]=\delta_{nm}$. The vacuum state of the corresponding Fock
space is annihilated by all $a_{n}$'s, i.e., $a_{n}\left|\,0\,\right\rangle=0$.

Before we construct our initial state let us briefly recall the characteristics of genuine multipartite
entanglement. Any pure quantum state that can be written as a tensor product with respect to any
bipartition is called \emph{bi-separable}. Generalizing this notion to mixed states, any mixed quantum
state is called bi-separable if it admits at least one decomposition into a convex sum of pure,
bi-separable states. Conversely, all states that are not bi-separable are called \emph{genuinely
multipartite entangled}. See, e.g., Refs.~\cite{GuehneToth2009,GabrielHiesmayrHuber2010} for more details.

We now proceed by constructing an initial state that contains no such genuine multipartite entanglement.
We select two of Rob's modes, $k$ and $k\pr$, and entangle them with modes $A$ and $C$ in Alice's and
Charlie's cavities respectively. This could be achieved by a scheme similar to that presented in
Ref.~\cite{BrownePlenio2003}. The initial, bi-separable state of the four modes $A,k,k\pr$ and $C$ we
denote by
\begin{align}
    \rho^{\,\pm}_{A\!RC}   &=  \ket{\!\Phi^{\pm}\!}\bra{\!\Phi^{\pm}\!}
    \label{eq:BosonsA 3 boxes initial density matrix}
\end{align}
with $\ket{\!\Phi^{\pm}\!}=\ket{\!\phi^{\pm}\!}_{A k}\ket{\!\phi^{\pm}\!}_{k\pr C}$, where the bipartite
entangled states are given by
\begin{subequations}
\label{eq:BosonsA 3 boxes initial states}
\begin{align}
    \left|\right.\!\phi^{\pm}\!\left.\right\rangle_{A k}  &=
    \tfrac{1}{\sqrt{2}}\,\bigl(\,\,\left|\,0\,\right\rangle\,
    \pm\,\left|\,1_{A}\,\right\rangle\,\left|\,1_{k}\,\right\rangle\,\bigr)\,,
    \label{eq:BosonsA 3 boxes initial state AR}\\[1.5mm]
    \left|\right.\!\phi^{\pm}\!\left.\right\rangle_{k\pr C}  &=
    \tfrac{1}{\sqrt{2}}\,\bigl(\,\,\left|\,0\,\right\rangle\,
    \pm\,\left|\,1_{k\pr}\,\right\rangle\,\left|\,1_{C}\,\right\rangle\,\bigr)\,.
    \label{eq:BosonsA 3 boxes initial state RC}
\end{align}
\end{subequations}
At $t=0$ we start to accelerate Rob's cavity uniformly and linearly and we let it follow a worldtube that
consists of segments of inertial motion and uniform acceleration. An example trajectory is shown in
Fig.~\ref{fig:multi cavity setup}. After an arbitrary number of such trajectory segments we assume without loss
of generality that the cavity remains inertial in the ``out-region''. The mode functions $\tilde{\phi}_{m}$ and
their corresponding annihilation and creation operators $\tilde{a}_{m}$ and $\tilde{a}^{\dagger}_{m}$ in the
out-region are related to their in-region counterparts by a Bogoliubov transformation with coefficients
$\alpha_{mn}$ and $\beta_{mn}$, i.e.,
\begin{align}
\tilde{\phi}_{m}    &=  \sum\limits_{n}\bigl(\alpha_{mn}\phi_{n}+\beta_{mn}\phi_{n}^{*}\bigr)
\label{eq:out region mode functions}
\end{align}
and
\begin{align}
\tilde{a}_{m}   &=  \sum\limits_{n}\bigl(\alpha^{*}_{mn}a_{n}-\beta^{*}_{mn}a^{\dagger}_{n}\bigr).
\label{eq:out region annihiliation operators}
\end{align}
In the case where the length
$\delta=(b-a)$ of the rigid cavity is fixed and the individual accelerations of the cavity at any point of the
trajectory are small compared to $1/\delta$ the Bogoliubov coefficients can be computed analytically as a Maclaurin
expansion, i.e.,
\begin{subequations}
\label{eq:boson bogo coefficients small h expansion}
\begin{align}
    \alpha_{mn} &=   \alpha^{_{(0)}}_{mn}+\alpha^{_{(1)}}_{mn}+O(h^{2})\,,
    \label{eq:alpha small h expansion}
    \\[2mm]
    \beta_{mn}  &=  \beta^{_{(1)}}_{mn}+\beta^{_{(2)}}_{mn}+O(h^{3})\,,
    \label{eq:beta small h expansion}
\end{align}
\end{subequations}
with $\alpha^{_{(0)}}_{mn}=\delta_{mn}G_{m}$ (no summation), see Ref.~\cite{BruschiFuentesLouko2012}. The $G_{m}$ are
phase factors satisfying $|G_{m}|^{2}=1$ that are picked up by the modes during the free time evolution in each segment
of the cavity motion. The superscripts $^{_{(n)}}$ in Eq.~(\ref{eq:boson bogo coefficients small h expansion}) indicate
the power of the expansion parameters $h_{i}$, which represent the products of the cavity width and the acceleration at
the center of the cavity in the $i$-th trajectory segment. For ease of notation we will suppress the subscript $i$ in the
remaining discussion.

The Bogoliubov coefficients for any trajectory that is composed of segments of inertial motion and uniform acceleration can
be constructed from the coefficients for a single switch from inertial motion to constant acceleration, i.e., coefficients
relating Minkowski modes $\phi_{n}$ and Rindler modes $\tilde{\phi}_{n}$, along with phases $G_{n}$ that are picked up by
the modes in each segment. A detailed derivation of these coefficients for massless and massive scalar fields as well the
construction of generic trajectories can be found in Refs.~\cite{BruschiFuentesLouko2012,BruschiDraganLeeFuentesLouko2012}.

Using well-known standard procedures (see, e.g., Refs.~\cite{BruschiFuentesLouko2012} or \cite{FriisBruschiLoukoFuentes2012})
the Fock states $\ket{\!n_{k}\!}$ and $\ket{\!n_{k\pr}\!}$ of Rob's cavity in the initial state $\rho^{\,\pm}_{A\!RC}$ can be
transformed to the out-region basis $\ket{\!\tilde{n}_{i}\!}$ to obtain the transformed density matrix
$\tilde{\rho}^{\,\pm}_{A\!RC}$. The transformations between the relevant in-region and out-region Fock states up to first
order in the perturbative expansion are given by (see Ref.~\cite{FriisBruschiLoukoFuentes2012})
\begin{subequations}
\label{eq:boson first order Fock state transformations}
\begin{align}
    \ket{0}\  &=\,  \ket{\tilde{0}}\,-\,\tfrac{1}{2}\sum\limits_{p,q}G_{q}^{\,*}\beta_{pq}^{_{(1)}\raisebox{-1.7pt}{\scriptsize{$*$}}}\,
                \tilde{a}^{\dagger}_{p}\tilde{a}^{\dagger}_{q}\,\ket{\tilde{0}}\,+\,O(h^{2})\,,
    \label{eq:boson vacuum first order transformation}\\[2mm]
    \ket{1_{k}} &=\,  G_{k}^{\,*}\,\ket{\!\tilde{1}_{k}\!}\,+\,\sum\limits_{m\neq k}
                    \alpha_{mk}^{_{(1)}\raisebox{-1.7pt}{\scriptsize{$*$}}}\,\ket{\!\tilde{1}_{m}\!}
                    \label{eq:boson single particle first order transformation}\\[1.5mm]
                &\, -\,\tfrac{1}{2}G_{k}^{\,*}\sum\limits_{p,q}G_{q}^{\,*}
                    \beta_{pq}^{_{(1)}\raisebox{-1.7pt}{\scriptsize{$*$}}}\,
                    \tilde{a}^{\dagger}_{p}\tilde{a}^{\dagger}_{q}\,\ket{\!\tilde{1}_{k}\!}\,+\,O(h^{2})\,,
                    \nonumber
\end{align}
\begin{align}
    \ket{\!1_{k}\!}\ket{\!1_{k\pr}\!}   &=\,    G_{k}^{\,*}\beta_{kk\pr}^{_{(1)}}\,\ket{\tilde{0}}\,+\,
                    G_{k}^{\,*}\!\sum\limits_{m\neq k\pr}
                    \alpha_{mk\pr}^{_{(1)}\raisebox{-1.7pt}{\scriptsize{$*$}}}\,
                    \tilde{a}_{m}^{\dagger}\,\ket{\!\tilde{1}_{k}\!}
                    \nonumber\\[1.5mm]
                &\, +\,G_{k}^{\,*}G_{k\pr}^{\,*}\,\ket{\!\tilde{1}_{k}\!}\ket{\!\tilde{1}_{k\pr}\!}\,+\,
                    G_{k\pr}^{\,*}\!\sum\limits_{m\neq k}\alpha_{mk}^{_{(1)}\raisebox{-1.7pt}{\scriptsize{$*$}}}\,\tilde{a}_{m}^{\dagger}\,\ket{\!\tilde{1}_{k\pr}\!}
                    \nonumber\\[1.5mm]
                &\, -\,\tfrac{1}{2}G_{k}^{\,*}G_{k\pr}^{\,*}\!\sum\limits_{p,q}G_{q}^{\,*}
                    \beta_{pq}^{_{(1)}\raisebox{-1.7pt}{\scriptsize{$*$}}}\,\tilde{a}_{p}^{\dagger}
                    \tilde{a}_{q}^{\dagger}\,\ket{\!\tilde{1}_{k}\!}\ket{\!\tilde{1}_{k\pr}\!}\,,
                    \label{eq:boson two particle first order transformation}
\end{align}
\end{subequations}
and the transformation of $\ket{\!1_{k\pr}\!}$ can be obtained from
(\ref{eq:boson single particle first order transformation}) by replacing $k$ with $k\pr$.

Subsequently, we trace over all of Rob's out-region modes except $k$ and $k\pr$, gaining the reduced density operator
$\tilde{\rho}^{\,\pm}_{Akk\pr C}$ of the four bosonic modes $A,k,k\pr$ and $C$. Keeping terms up to first order in $h$
we find that $\tilde{\rho}^{\,\pm}_{Akk\pr C}$ is effectively a state of two qubits, the modes $A$ and $C$, and two
qutrits, the modes $k$ and $k\pr$. In other words, to first order in $h$ the modes $A$ and $C$ are not further populated
by the Bogoliubov transformation and their respective Hilbert spaces can be truncated to single-qubit, i.e.,
two-dimensional, Hilbert spaces with basis vectors $\ket{\tilde{0}_{k}\!},\ket{\!\tilde{1}_{k}\!}$ and
$\ket{\tilde{0}_{k\pr}\!},\ket{\!\tilde{1}_{k\pr}\!}$ respectively. For the modes $k$ and $k\pr$, on the other hand, the
states $\ket{\!\tilde{2}_{k}\!}$ and $\ket{\!\tilde{2}_{k\pr}\!}$ are populated by the cavity motion. This means the
description of these modes involves at least three basis vectors each, i.e., the modes effectively become qutrits.

For such states a general quantification of genuine multipartite entanglement proves to be cumbersome, as there is no
general classification scheme for multipartite systems beyond qubits. However, we can construct an inequality that acts
as a witness for genuine multipartite entanglement by comparing diagonal- and off-diagonal elements of
$\tilde{\rho}^{\,\pm}_{Akk\pr C}$. The construction of such a witness is straightforward: We exploit permutation symmetries
of bi-separable pure states to construct a convex function of density matrix elements that satisfies a simple inequality.
The convexity of this function ensures that bi-separable mixed states satisfy the inequality. Consequently, its violation
detects genuine multipartite entanglement in mixed states, see Refs.~\cite{GabrielHiesmayrHuber2010,
MaChenChenSpenglerGabrielHuber2011,WuKampermannBruszKloecklHuber2012}. Since the diagonal elements that are newly generated
by the Bogoliubov transformation are at least $O(h^{2})$ the witness inequality, whose complete form is given by
Eq.~(\ref{eq:BosonsA GME witness complete}) of the appendix, takes the simple form
\begin{small}
\begin{align}
    2\,\left|
     \bra{\!1_{C}\!}
     \bra{\!\tilde{2}_{k\pr}\!}
     \bra{\!\tilde{2}_{k}\!}
     \bra{\!1_{A}\!}\,
     \tilde{\rho}^{\,\pm}_{Akk\pr C}\,
     \ket{\tilde{0}}\right|\,-\,O(h^{2})\,   &\leq\,0\,.
    \label{eq:BosonsA GME witness}
\end{align}
\end{small}
Evaluating the matrix element we can recast (\ref{eq:BosonsA GME witness}) as
\begin{align}
     \tfrac{1}{2}|\beta^{_{(1)}}_{kk\pr}|\,-\,O(h^{2})   &\leq\,0\,.
    \label{eq:BosonsA GME witness rewritten}
\end{align}
Within the perturbative regime this inequality is violated, i.e., genuine multipartite entanglement
is detected, whenever $\beta^{_{(1)}}_{kk\pr}\neq0$. This is the case for all mode pairs of opposite
parity, i.e., if $(k-k\pr)$ is odd. If the modes $k$ and $k\pr$ have the same parity the first order
coefficients relating these two modes vanish identically and statements about the violation of
inequality (\ref{eq:BosonsA GME witness}) can only be made for given mode numbers and the answers will
depend on the particular cavity motion. We will therefore employ (\ref{eq:BosonsA GME witness}) to first
order in $h$ as witness for the detection of genuine multipartite entanglement.

In fact, (\ref{eq:BosonsA GME witness rewritten}) is not only a witness, but also a lower bound to a
measure of genuine multipartite entanglement, the generalization of the concurrence to multipartite
entanglement. This type of concurrence is based on the linear entropy, which, in turn, is chosen in this
context only to write the witness in a compact form. However, any lower bound on a measure based on the
linear entropy supplies a lower bound to that same measure based on the physically more intuitive
R\'{e}nyi-2 entropy~\cite{WuKampermannBruszKloecklHuber2012}. The R\'{e}nyi-2 entropy can finally also be
related to the average minimal mutual information across any bipartition, minimized over all possible
decompositions. Moreover, to first order in the Maclaurin expansion $\rho^{\mathcal{B}}_{ARC}$ is a pure
state for which the bound is tight, as shown in Ref.~\cite{MaChenChenSpenglerGabrielHuber2011}.

In this respect the bounds provided by the witness inequality offer an immense advantage with respect to
the direct computation (if possible) of entropy-based measures, which require the perturbative corrections
to be calculated at least to second order in $h$, see Ref.~\cite{BruschiDraganLeeFuentesLouko2012}. The
reason for this lies in the dependence of entropic measures on the quantification of the mixedness of the
subsystems. The introduction of mixedness due to the Bogoliubov coefficients occurs only at second order of
the perturbative expansion, which renders entropy-based measures incapable of detecting such changes to
first order in $h$.

Conceptually the creation of multipartite correlations in the scenario with three, initially pairwise,
bipartite entangled cavities, could be interpreted in the following way: The non-uniform motion correlates
the non-interacting modes $k$ and $k\pr$ in Rob's cavity, such that all bi-partitions of the four-mode system
become entangled. As expected, the reduced two-mode state of Alice's and Charlie's cavities is unaffected by
the motion of Rob's cavity and, consequently, the modes $A$ and $C$ are still uncorrelated in the
out-region. The reduced state of the modes $k$ and $k\pr$ on the other hand becomes entangled, i.e., to
first order in $h$ the negativity $\mathcal{N}$ of
$\tilde{\rho}^{\,\pm}_{kk\pr}=\tr_{A,C}(\tilde{\rho}^{\,\pm}_{Akk\pr C})$ is given by
$\mathcal{N}=|\beta^{_{(1)}}_{kk\pr}|/4$. The negativity
\begin{align}
\mathcal{N} &=  \sum\limits_{i}\frac{|\lambda_{i}|-\lambda_{i}}{2}\,,
\label{eq:negativity definition}
\end{align}
which captures how the eigenvalues $\lambda_{i}$ of the partially transposed density matrix fail to be
positive, see Ref.~\cite{VidalWerner2001}, is a useful measure in the context of the perturbative
calculations at hand. It allows the quantification of bipartite entanglement at leading order, while entropic measures
of entanglement rely on the quantification of mixedness,
which is not altered to first order in $h$ and thus require the perturbative calculations to be extended to
higher orders. The fact that both the genuine multipartite entanglement of
$\tilde{\rho}^{\,\pm}_{Akk\pr C}$ and the bipartite entanglement of $\tilde{\rho}^{\,\pm}_{kk\pr}$ are
controlled by $\beta^{_{(1)}}_{kk\pr}$ supports the interpretation above.
However, we shall see that this naive view does not hold for the corresponding fermionic scenario
in Sec.~\ref{subsec:scenario A fermions}.

\begin{figure}[ht]
\centering
\vspace*{-4ex}%
\includegraphics[width=0.45\textwidth]{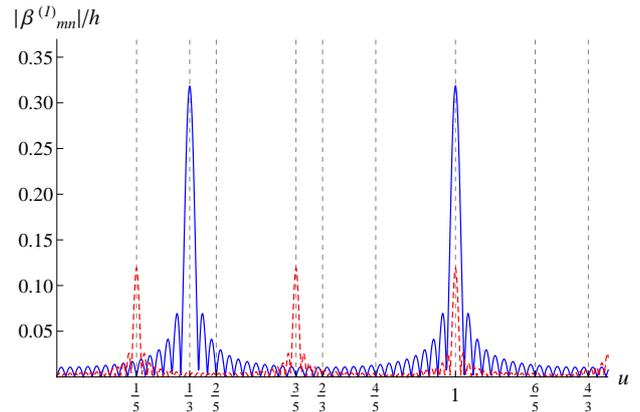}
\caption{
Illustration of the bosonic resonances:
The linear coefficients $|\beta^{_{(1)}}_{mn}|$ for the bosonic Bogoliubov transformations,
are shown for a real, massless scalar field in $(1+1)$ dimensions.
The travel scenario has $N$ segments of uniform acceleration $h/\delta$ and duration $\tau/2$
as measured at the cavity's centre,
separated by $(N-1)$ segments of inertial coasting of the same duration (to first order in $h$),
as illustrated in Fig.~\ref{fig:multi cavity setup} for $N=2$.
The curves in Fig.~\ref{fig:boson double resonance} are plotted for $N=15$,
$(m,n)=(k,k\pr)=(1,2)$ (blue,solid) and $(m,n)=(k\pr,k\prpr)=(2,3)$
(red,dashed) and $u:=h\tau/[4\delta\atanh(h/2)]$.
The vertical dashed lines indicate the potential resonance times for $(k,k\pr)$ and
$(k\pr,k\prpr)$ respectively. The explicit form of the Bogoliubov coefficients can be found in
Refs.~\cite{BruschiFuentesLouko2012,BruschiDraganLeeFuentesLouko2012}.
\label{fig:boson double resonance}}
\end{figure}

Nonetheless, the entanglement of the four-mode system is genuinely multipartite, which demonstrates
once again that genuine multipartite entanglement provides a richer structure than simple combinations
of bipartite correlations. Moreover, the responsible coefficient $|\beta^{_{(1)}}_{kk\pr}|$ can be
resonantly enhanced, within the limitations of the perturbative regime, by appropriate travel scenarios
when a massless scalar field is contained within the cavity~\cite{BruschiDraganLeeFuentesLouko2012}.

Such \emph{resonances} can occur if the chosen travel scenario consists of $N$ identical building blocks
of two or more different, inertial or uniformly-accelerated, trajectory segments. If the overall proper
time $\tau$ of one such building block, measured at the centre of the box, satisfies the
necessary resonance condition
\begin{align}
    \tau    &=  \tfrac{2\,n\,\delta}{k\,+\,k\pr}\,,
    \label{eq:resonance condition}
\end{align}
where $n$ is a positive integer,
and the coefficient $(\beta_{1})^{_{(1)}}_{kk\pr}$ of a single such building block is nonzero,
then the corresponding coefficient after $N$ repetitions scales linearly with $N$, i.e.,
\begin{align}
(\beta_{N})^{_{(1)}}_{kk\pr}    &=  N(\beta_{1})^{_{(1)}}_{kk\pr}\,.
\label{eq:resonance beta N shakes}
\end{align}
This scaling is valid within the perturbative regime, i.e., as long as $N h\ll 1$.
A detailed derivation of the condition in Eq.(\ref{eq:resonance condition}) can be found in
Ref.~\cite{BruschiDraganLeeFuentesLouko2012} but the discussion of the continuous variable techniques
necessary for the proof lies beyond the scope of this paper. See Fig.~\ref{fig:boson double resonance}
for an illustration of the resonances for different mode pairs.

The physical reason for the occurrence of these resonances lies in the
phases $G_{m}$ that are acquired by the modes during the free time evolution in the travel segments.
The Bogoliubov transformations that switch between segments of inertial and uniformly-accelerated
motion act as pairwise two-mode squeezing operations on all modes of opposite parity, see
Ref.~\cite{FriisFuentes2012,BruschiDraganLeeFuentesLouko2012}. When at least two such switches are
combined to a building block, the magnitude of the corresponding overall squeezing parameter depends
on the proper time between these switches as measured at the centre of the cavity. The quantities
$G_{m}$ are functions of this time and the mode numbers $m$. In particular, if, and only if, the quantum
field in question is massless, all frequencies are integer multiples of some basic frequency and the
phases can combine constructively to facilitate the resonance.

\subsection{Dirac fields}\label{subsec:scenario A fermions}

Let us now consider the analogous scenario for cavities containing fermionic rather than bosonic
quantum fields. In particular, let us assume that Alice's, Rob's and Charlie's cavities confine
Dirac fields, as discussed in Refs.~\cite{FriisLeeBruschiLouko2012,FriisBruschiLoukoFuentes2012}.
In the in-region the field in Rob's cavity satisfies the Dirac equation
\mbox{$(i\gamma^{\mu}\partial_{\mu}-m)\psi=0$} and the mode functions obey the boundary conditions
\begin{align}
(\psi^{\,\dagger}_{\omega}\gamma^{0}\gamma^{1}\psi_{\omega'})_{x=a} \,=\,
(\psi^{\,\dagger}_{\omega}\gamma^{0}\gamma^{1}\psi_{\omega'})_{x=b} &=  0\,.
\label{eq:fermion boundary conditions}
\end{align}
Here $\gamma^{\mu}$ $(\mu=0,1,2,3)$ are the usual Dirac matrices satisfying the anti-commutation relations
$\{\gamma^{\mu},\gamma^{\nu}\}=2\eta^{\mu\nu}$. Due to the boundary conditions the field modes
can be labelled by an integer $n$ and the field can be decomposed as
\begin{align}
\psi=\sum_{n\geq0}b_{n}\psi_{n}+\sum_{n<0}c_{n}^{\dagger}\psi_{n}\,.
\label{eq:in region Dirac field}
\end{align}
The operators $b_{n}$ and $c_{n}$
annihilate particles and antiparticles respectively. The vacuum state of the cavity is defined by
$b_{n}\fket{0}=c_{n}\fket{0}=0\ \forall n$, while the single particle
and antiparticle states, $\fket{\!1_{m}\!}^{\!\!+}$ and $\fket{\!1_{n}\!}^{\!\!-}$, are created from the vacuum by
$b_{m}^{\dagger}$ and $c_{n}^{\dagger}$ respectively. The fermionic operators further satisfy
$\{b_{m},b_{n}^{\dagger}\}=\{c_{m},c_{n}^{\dagger}\}=\delta_{mn}$, while all other anti-commutators vanish.

In analogy to the bosonic states (\ref{eq:BosonsA 3 boxes initial states}) we select initially bipartite
entangled fermionic states that correlate the cavities of Alice and Rob, as well as Rob and Charlie
respectively, i.e.,
\begin{subequations}
\label{eq:FermionsA 3 boxes initial states}
\begin{align}
    \fket{\phi^{\pm}}_{AR}   &=
    \tfrac{1}{\sqrt{2}}\,\bigl(\,\fket{0}\,
    \pm\,   \fket{\!1_{A}\!}^{\!\!-}\,
             \fket{\!1_{\kappa}\!}^{\!\!+}\,\bigr)\,,
    \label{eq:FermionsA 3 boxes initial state AR}\\[1.5mm]
    \fket{\phi^{\pm}}_{RC}   &=
    \tfrac{1}{\sqrt{2}}\,\bigl(\,\fket{0}\,
    \pm\,   \fket{\!1_{\kappa\pr}\!}^{\!\!-}\,
             \fket{\!1_{C}\!}^{\!\!+}\,\bigr)\,,
    \label{eq:FermionsA 3 boxes initial state RC}
\end{align}
\end{subequations}
where we have chosen particle and antiparticle modes in agreement with charge superselection rules.
The multi-particle states in Eq.~(\ref{eq:FermionsA 3 boxes initial states}) are elements of
anti-symmetrized tensor product spaces of single particle mode functions $\psi_{i}$, i.e.,
\begin{align}
    \fket{\!1_{p}\!}^{\!\!+}\,\fket{\!1_{q}\!}^{\!\!-}  &:=
    \fket{\!\psi_{p}\!}^{\!\!+}\!\wedge\fket{\!\psi_{q}\!}^{\!\!-}\,:=\,
    b^{\dagger}_{p}c^{\dagger}_{q}\,\fket{0}\,,
    \label{eq:fermion multiparticle state convention}
\end{align}
\vspace*{-0.5mm}
where $A\wedge B=(A\otimes B-B\otimes A)/\sqrt{2}$ denotes the anti-symmetrized tensor product.
For the purpose of quantum information procedures one can map this ``wedge product'' structure
to the usual tensor product structure by defining a convention for partial tracing. Our choice
here is to trace out operators ``from the inside'', i.e.,
\begin{align}
    \tr_{p}\Bigl(b^{\dagger}_{\kappa}b^{\dagger}_{p}\,\fket{0}\fbra{0}\,b_{p}\Bigr) &=\,
    b^{\dagger}_{\kappa}\,\fket{0}\fbra{0}\,.
    \label{eq:fermion multiparticle trace convention}
\end{align}
\vspace*{-0.5mm}
In the spirit of this convention we reverse the ordering of the states in the adjoint space, i.e.,
$\fbra{0}b_{p}b_{\kappa}=^{\,+\!\!\!}\fbra{\!1_{p}\!}\!^{\,+\!\!\!}\fbra{\!1_{\kappa}\!}$. Different
conventions can generally introduce ambiguities in the resulting fermionic states, which has been the
subject of an on-going debate, see Ref.~\cite{AmbiguityDebate}.
However, we have verified that no ambiguities appear in our results to second order in the perturbative
calculations. For the symmetrized tensor product structure of bosonic modes analogous mappings have to
be performed. However, no sign ambiguities occur for bosons and the results are therefore independent of
the chosen mapping.

The Bogoliubov transformation that relates the in-region modes $\psi_{n}$ and the out-region modes
$\tilde{\psi}_{m}$ is given by
\vspace*{-1.5mm}
\begin{align}
\tilde{\psi}_{m}    &=  \sum\limits_{n}\mathcal{A}_{mn}\psi_{n}
\label{eq:fermionic in to out region bogos}
\end{align}
where $n\in\mathds{Z}$ and
the fermionic Bogoliubov coefficients can be expanded in a Maclaurin series in the parameter $h$ as
\vspace*{-1.5mm}
\begin{align}
    \mathcal{A}_{mn}    &=\,\mathcal{A}_{mn}^{_{(0)}}+\mathcal{A}_{mn}^{_{(1)}}+O(h^{2})\,,
    \label{eq:fermionic bogo coefficient expansion}
\end{align}
where the superscript $^{_{(n)}}$ again denotes the power of $h$, and we have $\mathcal{A}^{_{(1)}}_{nn}=0$ and
$\mathcal{A}^{_{(0)}}_{mn}=\delta_{mn}G_{m}$~(no summation). In
the fermionic case the subscripts $m$ and $n$ can take positive as well as negative values, representing particle and
antiparticle modes respectively. Furthermore, the splitting into particle and antiparticle modes is controlled by an
additional parameter in the Bogoliubov coefficients, see Ref.~\cite{FriisLeeBruschiLouko2012}. We set this parameter
to zero, in which case the phase factors $G_{m}$ coincide for bosons and fermions.

We continue, as previously, by transforming the initial state
$\fket{\!\Phi^{\pm}\!}=\fket{\!\phi^{\pm}\!}_{A\kappa}\fket{\!\phi^{\pm}\!}_{\kappa\pr C}$
with the corresponding density operator $\varrho^{\,\pm}_{A\!RC}=\fket{\!\Phi^{\pm}\!}\fbra{\!\Phi^{\pm}\!}$,
to the out-region states, i.e., we use the series expansions~\cite{FriisLeeBruschiLouko2012}
\begin{subequations}
\label{eq:fermionic in region states transformed}
\begin{align}
\hspace*{-1.0mm}
\fket{0}\ \;   &=  \fket{\tilde{0}}\,+\,
    \sum_{p,q}\,G_{q}^{\,*}\mathcal{A}^{_{(1)}\raisebox{-1.7pt}{\scriptsize{$*$}}}_{pq}\,
    \tilde{b}_{p}^{\dagger}\tilde{c}_{q}^{\dagger}\,\fket{\tilde{0}}\,+\,O(h^{2})\,,
    \label{eq:fermionic vacuum in region transformation}\\[0.5mm]
\hspace*{-1.0mm}
\fket{\!1_{\kappa}\!}^{\!\!+}\,    &=  G^{\,*}_{\kappa}\fket{\!\tilde{1}_{\kappa}\!}^{\!\!+}\,+\,
    \sum_{m\geq0}\,\mathcal{A}^{_{(1)}\raisebox{-1.7pt}{\scriptsize{$*$}}}_{m\kappa}\,
    \,\fket{\!\tilde{1}_{m}\!}^{\!\!+}
    \label{eq:fermionic particle in region transformation}\\
    &\ \ +\,G_{\kappa}^{\,*}\sum\limits_{p,q}\,
    G_{q}^{\,*}\mathcal{A}^{_{(1)}\raisebox{-1.7pt}{\scriptsize{$*$}}}_{pq}\,
    \tilde{b}_{p}^{\dagger}\tilde{c}_{q}^{\dagger}\,\fket{\!\tilde{1}_{\kappa}\!}^{\!\!+}
    \,+\,O(h^{2})\,,\nonumber\\[0.5mm]
\hspace*{-1.0mm}
\fket{\!1_{\kappa\pr}\!}^{\!\!-}    &=  G^{}_{\kappa\pr}\fket{\!\tilde{1}_{\kappa\pr}\!}^{\!\!-}\,+\,
    \sum_{n<0}\,\mathcal{A}^{_{(1)}}_{n\kappa\pr}\,
    \,\fket{\!\tilde{1}_{n}\!}^{\!\!-}
    \label{eq:fermionic antiparticle in region transformation}\\
    &\ \ +\,G_{\kappa\pr}^{}\sum\limits_{p,q}\,
    G_{q}^{\,*}\mathcal{A}^{_{(1)}\raisebox{-1.7pt}{\scriptsize{$*$}}}_{pq}\,
    \tilde{b}_{p}^{\dagger}\tilde{c}_{q}^{\dagger}\,\fket{\!\tilde{1}_{\kappa\pr}\!}^{\!\!-}
    \,+\,O(h^{2})\,,\nonumber
\end{align}
\begin{align}
\hspace*{-1.5mm}
\fket{\!1_{\kappa}\!}^{\!\!+}\!\fket{\!1_{\kappa\pr}\!}^{\!\!-}   &\,=
    G_{\kappa\pr}G_{\kappa}^{\,*}
    \fket{\!\tilde{1}_{\kappa}\!}^{\!\!+}\!\fket{\!\tilde{1}_{\kappa\pr}\!}^{\!\!-}\,+\,
    G_{\kappa\pr}\mathcal{A}^{_{(1)}\raisebox{-1.7pt}{\scriptsize{$*$}}}_{\kappa\pr\kappa}\,\fket{\tilde{0}}
    \nonumber\\[2.0mm]
    &\ +\,G_{\kappa\pr}G_{\kappa}^{\,*}
    \sum_{p,q}\,G_{q}^{\,*}\mathcal{A}^{_{(1)}\raisebox{-1.7pt}{\scriptsize{$*$}}}_{pq}\,
    \tilde{b}_{p}^{\dagger}\tilde{c}_{q}^{\dagger}\,
    \fket{\!\tilde{1}_{\kappa}\!}^{\!\!+}\!\fket{\!\tilde{1}_{\kappa\pr}\!}^{\!\!-}
    \nonumber\\
    &\ +\,G_{\kappa\pr}\sum\limits_{m\geq0}\mathcal{A}^{_{(1)}\raisebox{-1.7pt}{\scriptsize{$*$}}}_{m\kappa\pr}\,
    \fket{\!\tilde{1}_{m}\!}^{\!\!+}\!\fket{\!\tilde{1}_{\kappa\pr}\!}^{\!\!-}
    \label{eq:fermionic particle antiparticle pair transformed}\\
    &\ +\,G_{\kappa}^{\,*}\sum\limits_{n<0}\mathcal{A}^{_{(1)}}_{n\kappa\pr}\,
    \fket{\!\tilde{1}_{\kappa}\!}^{\!\!+}\!\fket{\!\tilde{1}_{n}\!}^{\!\!-}\,+\,O(h^{2})\,,
    \nonumber
\end{align}
\end{subequations}
where we have used the convention from Eqs.~(\ref{eq:fermion multiparticle state convention},\ref{eq:fermion multiparticle
trace convention}), to obtain the transformed state $\tilde{\varrho}^{\,\pm}_{A\!RC}$.

Subsequently we trace over all particle and antiparticle modes except $\kappa\geq0$ and $\kappa\pr<0$, which leaves
us with the reduced state $\tilde{\varrho}^{\,\pm}_{A\kappa\kappa\pr C}$.
To first order in the parameter $h$ no new diagonal elements are generated in the state transformation, which makes
it easy to identify an inequality that acts as a witness for genuine multipartite entanglement. However, due to the
Pauli exclusion principle, we cannot employ a witnesses of the type of Eq.~(\ref{eq:BosonsA GME witness}) for the
fermionic system. Instead we construct the witness
\begin{align}
    &\left|^{\raisebox{0.7pt}{\scriptsize{$-$}}\!\!\!}\fbra{\!\tilde{1}_{\kappa\pr}\!}
     ^{\,+\!\!\!}\fbra{\!\tilde{1}_{\kappa}\!}
     \tilde{\varrho}^{\,\pm}_{A\kappa\kappa\pr C}
     \fket{\!\tilde{1}_{\kappa\pr}\!}^{\!\!-}
     \fket{\!1_{C}\!}^{\!\!+}\right|
     \label{eq:FermionsA GME witness}\\[1.5mm]
     \,+\,
     &\left|^{\raisebox{0.7pt}{\scriptsize{$+$}}\!\!\!}\fbra{\!1_{C}\!}
      ^{-\!\!\!}\fbra{\!1_{A}\!}
     \tilde{\varrho}^{\,\pm}_{A\kappa\kappa\pr C}
     \fket{\!\tilde{1}_{\kappa\pr}\!}^{\!\!-}
     \fket{\!1_{C}\!}^{\!\!+}\right|
     \,-\,O(h^{2})\,\leq   0\,,
    \nonumber
\end{align}
which we can express as
\begin{align}
    \tfrac{1}{2}|\mathcal{A}^{_{(1)}}_{\kappa\kappa\pr}|
    \,-\,O(h^{2})\,&\leq   0\,.
    \label{eq:FermionsA GME witness rewritten}
\end{align}
The complete form of the witness can be found in the appendix, see Eq.~(\ref{eq:FermionsA GME witness complete}).
As with its bosonic counterpart~(\ref{eq:BosonsA GME witness rewritten}) the
inequality~(\ref{eq:FermionsA GME witness rewritten}) is always violated if the
modes $\kappa$ and $\kappa\pr$ have opposite parity, in which case
$\mathcal{A}^{_{(1)}}_{\kappa\kappa\pr}\neq0$. In the case where \mbox{$(\kappa+\kappa\pr)$} is even
the first order coefficient vanishes identically, see Ref.~\cite{FriisLeeBruschiLouko2012},
and the usefulness of the witness has to be evaluated for each selection of mode numbers and
travel scenarios individually.

The witness employed in Eq.~(\ref{eq:FermionsA GME witness rewritten}) is again a lower bound to the minimal average
mutual information across any bipartition~\cite{WuKampermannBruszKloecklHuber2012}, although, this time, it is not
tight for pure states as Eq.~(\ref{eq:BosonsA GME witness}).
However, there are other conceptually intriguing features appearing
for the fermionic four-qubit state $\tilde{\varrho}^{\,\pm}_{A\kappa\kappa\pr C}$. First, we notice that the
unperturbed reduced density operators $\varrho^{\,\pm}_{AC}$ and $\varrho^{\,\pm}_{\kappa\kappa\pr}$ are both
maximally mixed. This means that, in contrast to the bosonic case, no bipartite entanglement between the fermionic
modes $\kappa$ and $\kappa\pr$ can be generated from the initial state $\varrho^{\,\pm}_{A\!RC}$ by small
perturbations. The negativity, which is nonzero for all
entangled two-qubit states, requires (at least) one of the degenerate eigenvalues $1/4$ of the
partial transpose to become negative. This cannot happen within the perturbative regime. This behaviour can be
readily understood in terms of the monogamy of entanglement (see, e.g., Ref.~\cite{OsborneVerstraete2006}): To
first order in the small parameter expansion the reduced states of the initially maximally entangled modes,
$A$ and $\kappa$ or $\kappa\pr$ and $C$, respectively, remain unperturbed up to relative phases due to the time
evolution. In particular, the bipartite entanglement between $A$ and $\kappa$, as well as between $\kappa\pr$ and
$C$ is maximal to first order in $h$, see Ref.~\cite{FriisLeeBruschiLouko2012}, which excludes the possibility of
first order correlations between $\kappa$ and $\kappa\pr$. This remarkable difference between the scalar field and
the Dirac field highlights once more (see, e.g., Refs.~\cite{AlsingFuentes-SchullerMannTessier2006,AdessoFuentes-SchullerEricsson2007,AdessoFuentes-Schuller2009})
the contrast of fermionic and bosonic particle statistics in the context of Bogoliubov transformations.

Furthermore, we can
straightforwardly classify the entanglement of the four qubit state. We notice that, to first order, the state
$\tilde{\varrho}^{\,\pm}_{A\kappa\kappa\pr C}$ can be decomposed as
\begin{align}
    \tilde{\varrho}^{\,\pm}_{A\kappa\kappa\pr C}    &=
    \fket{\!\mathcal{D}\!}\fbra{\!\mathcal{D}\!}
    \,+\,O(h^{2})
    \label{eq:FermionsA Dicke state decomposition}
\end{align}
where $\left|\!\left|\right.\right.\!\mathcal{D}\!\left.\left.\right\rangle\!\right\rangle$ is a \emph{Dicke
state}~\cite{Dicke1954,HuberErkerSchimpfGabrielHiesmayr2011}, given by
\begin{align}
     \fket{\!\mathcal{D}\!}
     &=
     \tfrac{1}{2}\bigl(\,
        \fket{\tilde{0}}
        \,\pm\,
        G^{\,*}_{\kappa}
        \fket{\!1_{A}\!}^{\!\!-}\!
        \fket{\!\tilde{1}_{\kappa}\!}^{\!\!+}
        \,\pm\,
        G_{\kappa\pr}
        \fket{\!\tilde{1}_{\kappa\pr}\!}^{\!\!-}\!
        \fket{\!1_{C}\!}^{\!\!+}
        \nonumber\\[1.5mm]
        &\ +\,
        G_{\kappa\pr}\mathcal{A}^{_{(1)}\raisebox{-1.7pt}{\scriptsize{$*$}}}_{\kappa\kappa\pr}
        \fket{\!\tilde{1}_{\kappa}\!}^{\!\!+}\!
        \fket{\!\tilde{1}_{\kappa\pr}\!}^{\!\!-}
        \,-\,
        G^{\,*}_{\kappa}\mathcal{A}^{_{(1)}}_{\kappa\kappa\pr}
        \fket{\!1_{A}\!}^{\!\!-}\!
        \fket{\!1_{C}\!}^{\!\!+}
        \nonumber\\[1.5mm]
        &\ -\,
        G^{\,*}_{\kappa}G_{\kappa\pr}
        \fket{\!1_{A}\!}^{\!\!-}\!
        \fket{\!\tilde{1}_{\kappa}\!}^{\!\!+}\!
        \fket{\!\tilde{1}_{\kappa\pr}\!}^{\!\!-}\!
        \fket{\!1_{C}\!}^{\!\!+}
             \,\bigr)\,,
    \label{eq:FermionsA Dicke state}
\end{align}
where $G_{m}$ are mode dependent phase factors of unit magnitude, i.e., $|G_{m}|=1$, that are determined
by the specific travel scenario, see, e.g., Ref.~\cite{FriisLeeBruschiLouko2012}. The usual form of the
Dicke state can be obtained from Eq.~(\ref{eq:FermionsA Dicke state decomposition}) by local unitaries, e.g.,
bit-flips in the modes $A$ and $\kappa\pr$.\\

\section{Single-cavity entanglement}\label{sec:scenario B}

Let us now modify our previous setup and consider only Rob's cavity on its own, as in
Ref.~\cite{FriisBruschiLoukoFuentes2012}. In particular, let us assume for simplicity that
the initial state of the in-region modes in that cavity is the vacuum. We want to investigate the
genuine multipartite correlations that might possibly be generated between three selected modes
by performing a Bogoliubov transformation to the out-region modes.\\

\subsection{Scalar fields}\label{subsec:scenario B bosons}

The in-region vacuum $\ket{0}$ of a real scalar field $\phi$ can be related to the out-region vacuum $\ket{\tilde{0}}$
by $\ket{0}=M e^{W}\ket{\tilde{0}}$, where $M$ is a normalization constant and
$W:=\tfrac{1}{2}\sum_{i,j}V_{ij}\tilde{a}_{i}^{\dagger}\tilde{a}_{j}^{\dagger}$. The coefficients $V_{ij}$ form a
symmetric matrix that can be expressed as $V=-\beta^{\,*}\alpha^{-1}$, where $\beta=(\beta_{mn})$ and it is implicitly
assumed that $\alpha=(\alpha_{mn})$ is invertible. We then apply the small parameter expansion for the Bogoliubov
coefficients and find $V=V^{\raisebox{0.7pt}{\tiny{$(1)$}}}+O(h^{2})$ and the normalization constant
$M=(1-\tfrac{1}{4}\sum_{i,j}|V_{ij}|^{2})+O(h^{3})$, see Ref.~\cite{FriisBruschiLoukoFuentes2012}.

Having transformed the initial vacuum state $\ket{0}$ to the out-region, we continue by tracing over all out-region
modes except three chosen modes $k$, $k\pr$ and $k\prpr$. We denote the transformed, reduced state by
$\tilde{\rho}_{k\,k\pr k\prpr}$, where we keep terms up to second order in $h$. At this stage we further assume that the
modes do not all have the same parity, e.g., let us choose $(k-k\pr)$ and $(k\pr-k\prpr)$ to be odd, which implies
$(k-k\prpr)$ is even. This further means that $|\,\beta^{_{(1)}}_{k\,k\pr}|\geq0$, $|\,\beta^{_{(1)}}_{k\pr k\prpr}|\geq0$, while
$|\,\beta^{_{(1)}}_{k\,k\prpr}|=0$. With this convention in mind we select a witness for genuine multipartite entanglement, i.e.,
\begin{align}
    2\,|\bra{\tilde{0}}\,\tilde{\rho}_{k\,k\pr k\prpr}\,\ket{\!\tilde{1}_{k}\!}\ket{\!\tilde{2}_{k\pr}\!}\ket{\!\tilde{1}_{k\prpr}\!}|
    \,-\,O(h^{3})   &\leq0\,,
    \label{eq:BosonsB GME witness}
\end{align}
which can be rewritten as
\begin{align}
    2\sqrt{2}\,|\,\beta^{_{(1)}}_{k\,k\pr}|\,|\,\beta^{_{(1)}}_{k\pr k\prpr}|
    \,-\,O(h^{3})   &\leq0\,.
    \label{eq:BosonsB GME witness rewritten}
\end{align}
As previously, the witness (\ref{eq:BosonsB GME witness}) presents a lower bound to
the convex roof extension of the minimal average mutual information across all
bi-partitions~\cite{WuKampermannBruszKloecklHuber2012} and its complete form is given by
(\ref{eq:BosonsB GME witness complete}) of the appendix. It can be immediately noticed that
that the previously discussed bosonic resonances~\cite{BruschiDraganLeeFuentesLouko2012}
allow the linear enhancement of the individual coefficients, $|\,\beta^{_{(1)}}_{k\,k\pr}|$ or
$|\,\beta^{_{(1)}}_{k\pr k\prpr}|$, for particular basic travel times $\tau=2n\delta/(k+k\pr)$
and $\tau=2m\delta/(k\pr+k\prpr)$, $n,m\in\mathds{N}_{\!+}$, respectively. Interestingly, these
resonances coincide for $n=p(k+k\pr)$ and $m=p(k\pr+k\prpr)$, i.e., when the travel time, as
measured at the centre of the cavity, for a single cycle of the repeated basic travel scenario
is $\tau=2p\delta$, $p\in\mathds{N}_{\!+}$, see Fig.~\ref{fig:boson double resonance}.

At this mode-independent resonance the lower bound on the genuine multipartite entanglement
increases quadratically with the number $N$ of repetitions of the basic travel block. At the
same time the mixedness that is introduced to the system by tracing out the other modes
contains second order terms $f^{\beta}_{k\lnot k\pr}$, $f^{\beta}_{k\pr\lnot k,k\prpr}$ and
$f^{\beta}_{k\prpr\lnot k\pr}$, where
$f^{\beta}_{m\lnot p}=\tfrac{1}{2}\sum_{n\neq p}|\,\beta^{_{(1)}}_{mn}|^{2}$, which all
exhibit quadratic scaling at the mode-independent resonance. The validity of the
perturbative approach is ensured since all second order terms are at most proportional to
$N^{2}h^{2}\ll N h$.

However, the classification of the bosonic genuine multipartite correlations remains an
unsolved problem. To first order the transformed state can be written as a pure Dicke
state~\cite{HuberErkerSchimpfGabrielHiesmayr2011}, but since the genuine multipartite
entanglement is detected at second order this is of no significance. To second order the
modes effectively become qutrits, for which generally little is known about entanglement
classes.\\

\subsection{Dirac fields}\label{subsec:scenario B fermions}

Let us again consider the fermionic counterpart to the bosonic situation. The in-region vacuum
$\fket{0}$ of the Dirac field $\psi$ is related to the out-region vacuum $\fket{\tilde{0}}$ by
$\fket{0}=\mathcal{M} e^{\mathcal{W}}\fket{\tilde{0}}$, where
$\mathcal{W}:=\sum_{\substack{\mbox{\tiny{$p\!\geq\!0$}}\\ \mbox{\tiny{$q\!<\!0$}}}}
\mathcal{V}_{pq}b^{\dagger}_{p}c^{\dagger}_{q}$ and $\mathcal{M}$ is a
normalization constant. Working perturbatively in the parameter $h$ the coefficient matrix
$\mathcal{V}$ can be expanded in a Maclaurin series as
$\mathcal{V}=\mathcal{V}^{\raisebox{0.7pt}{\tiny{$(1)$}}}+O(h^{2})$
and $\mathcal{V}^{_{(1)}}_{pq}=G_{q}\mathcal{A}^{_{(1)}\raisebox{-1.7pt}{\scriptsize{$*$}}}_{pq}$
(no summation), where the $G_{m}$ are mode specific phase factors, i.e., $|G_{m}|=1$, that depend
on the chosen travel scenario, see
Ref.~\cite{FriisLeeBruschiLouko2012}. The normalization constant can be found to be
$\mathcal{M}=1-\tfrac{1}{2}\sum_{\substack{\mbox{\tiny{$p\!\geq\!0$}}\\ \mbox{\tiny{$q\!<\!0$}}}}
|\mathcal{V}^{_{(1)}}_{pq}|^{2}+O(h^{3})$.

We can then perform the Bogoliubov transformation of the in-region vacuum $\fket{0}$ to the
out-region state and trace over all modes except three chosen modes $\kappa\geq0$, $\kappa\pr\geq0$
and $\kappa\prpr<0$, obtaining the state $\tilde{\varrho}_{\kappa\,\kappa\pr\kappa\prpr}$.
Since we do not expect any coherence effects between modes of the same charge
when we start from the vacuum, see Ref.~\cite{FriisBruschiLoukoFuentes2012}, we further assume
that $(\kappa+\kappa\pr)$ is even, while $(\kappa+\kappa\prpr)$ and $(\kappa\pr+\kappa\prpr)$
are odd. Specializing to the massless case this implies that
$|\mathcal{A}^{_{(1)}}_{\kappa\kappa\prpr}|\geq0$ and
$|\mathcal{A}^{_{(1)}}_{\kappa\pr\kappa\prpr}|\geq0$, while
$|\mathcal{A}^{_{(1)}}_{\kappa\kappa\pr}|=0$.
\vspace*{0.5mm}

For the state $\tilde{\varrho}_{\kappa\,\kappa\pr\kappa\prpr}$ we can form a witness for genuine
multipartite entanglement using the techniques from Ref.~\cite{HuberMintertGabrielHiesmayr2010}.
The violation of the inequality
\begin{small}
\begin{align}
&   |\fbra{\tilde{0}}
    \tilde{\varrho}_{\kappa\,\kappa\pr\kappa\prpr}
    \fket{\!\tilde{1}_{\kappa}\!}^{\!\!+}\fket{\!\tilde{1}_{\kappa\prpr}\!}^{\!\!-}|+
    |\fbra{\tilde{0}}
    \tilde{\varrho}_{\kappa\,\kappa\pr\kappa\prpr}
    \fket{\!\tilde{1}_{\kappa\pr}\!}^{\!\!+}\fket{\!\tilde{1}_{\kappa\prpr}\!}^{\!\!-}|
    \nonumber\\[1.5mm]
&   -\Bigl[
    \fbra{\tilde{0}}
    \tilde{\varrho}_{\kappa\,\kappa\pr\kappa\prpr}
    \fket{\tilde{0}}
    \Bigl(
    ^{\raisebox{-3.3pt}{\tiny{$-$}}\!\!}\fbra{\!\tilde{1}_{\kappa\prpr}\!}^{\!+\!\!}\fbra{\!\tilde{1}_{\kappa}\!}
    \tilde{\varrho}_{\kappa\,\kappa\pr\kappa\prpr}
    \fket{\!\tilde{1}_{\kappa}\!}^{\!\!+}\fket{\!\tilde{1}_{\kappa\prpr}\!}^{\!\!-}
    \nonumber\\[-0.5mm]
&   +
    ^{-\!\!\!\!}\fbra{\!\tilde{1}_{\kappa\prpr}\!}^{+\!\!\!}\fbra{\!\tilde{1}_{\kappa\pr}\!}
    \tilde{\varrho}_{\kappa\,\kappa\pr\kappa\prpr}
    \fket{\!\tilde{1}_{\kappa\pr}\!}^{\!\!+}\fket{\!\tilde{1}_{\kappa\prpr}\!}^{\!\!-}
    \Bigr)\Bigr]^{\raisebox{-0.5pt}{\tiny{$\tfrac{1}{2}$}}}
    -O(h^{2})\leq 0
\label{eq:FermionsB witness}
\end{align}
\end{small}
detects genuine multipartite entanglement. The complete form without perturbative expansion is given by
Eq.~(\ref{eq:FermionsB witness complete}) in the appendix. It can be expressed in
the simple form
\begin{small}
\begin{align}
    |\mathcal{A}^{_{(1)}}_{\kappa\kappa\prpr}|+
    |\mathcal{A}^{_{(1)}}_{\kappa\pr\kappa\prpr}|-
    \sqrt{|\mathcal{A}^{_{(1)}}_{\kappa\kappa\prpr}|^{2}+
    |\mathcal{A}^{_{(1)}}_{\kappa\pr\kappa\prpr}|^{2}}+O(h^{2})\leq0\,.
    \label{eq:FermionsB witness rewritten}
\end{align}
\end{small}
Using the triangle inequality this can be easily seen to be violated whenever both
$\mathcal{A}^{_{(1)}}_{\kappa\kappa\prpr}$ and $\mathcal{A}^{_{(1)}}_{\kappa\pr\kappa\prpr}$
are nonzero.

We can further classify the genuine multipartite entanglement in this case, since the state
$\tilde{\varrho}_{\kappa\,\kappa\pr\kappa\prpr}$ admits the decomposition
\begin{align}
\tilde{\varrho}_{\kappa\,\kappa\pr\kappa\prpr}
    &=\,\fket{\!\mathrm{W}\!}\fbra{\!\mathrm{W}\!}+O(h^{2})
\label{eq:par-tr not k kprime kprimeprime vacuum}
\end{align}
where the class-defining $W$-state is
\begin{align}
 \fket{\!\mathrm{W}\!}
&=  \,\fket{\tilde{0}}\,
    +\,G_{\kappa\prpr}\mathcal{A}_{\kappa\pr\kappa\prpr}^{_{(1)}\raisebox{-1.7pt}{\scriptsize{$*$}}}\,
    \fket{\!\tilde{1}_{\kappa\pr}\!}^{\!\!+}\fket{\!\tilde{1}_{\kappa\prpr}\!}^{\!\!-}
    \nonumber\\[1.5mm]
&\ \, +\,G_{\kappa\prpr}\mathcal{A}_{\kappa\kappa\prpr}^{_{(1)}\raisebox{-1.7pt}{\scriptsize{$*$}}}\,
    \fket{\!\tilde{1}_{\kappa}\!}^{\!\!+}\fket{\!\tilde{1}_{\kappa\prpr}\!}^{\!\!-}\,.
\label{eq:W state}
\end{align}

\section{Conclusions}\label{sec:conclusions}

We have studied genuine multipartite entanglement of bosonic and fermionic modes of
relativistic quantum fields in non-uniformly moving cavities. We have used the perturbative
approach of Refs.~\cite{BruschiFuentesLouko2012,FriisLeeBruschiLouko2012} to handle the
Bogoliubov transformations that feature in the transformation of the cavity modes between
the inertial in-region and out-region. The non-uniform motion in between these regions
populates modes and shifts pre-existing excitations. The final out-region states of a set
of chosen modes is then obtained by tracing over all other modes.
In two qualitatively different scenarios we have shown that genuine multipartite
correlations are generated from initially bi-separable or separable states of the chosen modes.

We have employed witnesses for multipartite entanglement that prove to be advantageous
with respect to usual entropic measures in the perturbative regime. The witnesses for
the bosonic correlations provide lower bounds to measures of genuine multipartite
entanglement that are based on convex roof extensions of the minimal average mutual
information over all bi-partitions~\cite{WuKampermannBruszKloecklHuber2012}.
We find that the numerical value of the perturbative corrections to these lower bounds can
be resonantly enhanced for any chosen triple of bosonic modes with varying parities.
However, the classification of the arising correlations is hindered by the unknown
classification structure beyond qubits.

For the fermionic systems, on the other hand, we can detect the genuine multipartite
entanglement in the transformed states and, subsequently, assign the entangled states
originating from initially bi-separable and separable states to the classes of 4-qubit
Dicke states, which can be used in quantum secret sharing~
\cite{GaertnerKurtsieferBourennaneWeinfurter2007}, and 3-qubit $W$-states
respectively.

The creation of specific entangled states in our setup can be considered as the
realization of quantum gates by motion in spacetime. In particular, we complement the findings
of Refs.~\cite{FriisFuentes2012,BruschiDraganLeeFuentesLouko2012}, where two-mode
squeezing gates are implemented as a result of the non-uniform cavity motion, with
the creation of the Dicke and $W$ states for fermionic systems.

Indeed, the cavity setups studied here and in
Refs.~\cite{FriisBruschiLoukoFuentes2012,FriisFuentes2012,BruschiDraganLeeFuentesLouko2012}
share intriguing features with the models for frequency combs, which are known to produce
cluster states, a vital resource for universal quantum computation~\cite{MenicucciFlammiaPfister2008}.
Future work is being directed towards the investigation of this connection as well as to the
extraction of the cavity mode entanglement with suitable detector models~\cite{BruschiFuentesKempfLeeLouko2012}.

Additionally, the presence of the genuine multipartite correlations in these relativistic
settings can be used for high precision tests of the quantumness of correlations and might have
significant advantages in identifying the signatures of quantum phenomena where
relativistic effects are notoriously small. The resonant behaviour of the bosonic
multipartite entanglement presented here can be a keystone countermeasure to this problem
and allows precise control and, hopefully, experimental testing of such effects.

Finally, we believe that these observations offer a fundamentally new viewpoint on Bogoliubov
transformations: The Bogoliubov coefficients are not mere indicators of average particle numbers,
they are responsible for genuine multi-mode coherence in genuinely-multipartite entangled quantum
systems.\\

\begin{acknowledgments}
\vspace*{-3mm}
We thank G.~Adesso, D.~Girolami, M.~I.~Gu\c{t}\u{a}, A.~Kempf, P.~Kok, A.~R. Lee, J.~Louko and
E.~Mart\'{i}n-Mart\'{i}nez for helpful discussions and comments.
We are grateful for support by the University of Bristol and the $\chi$-QEN collaboration.
N.~F. and I.~F. acknowledges support from EPSRC (CAF Grant No.~EP/G00496X/2 to I.~F).
\end{acknowledgments}

\vspace*{-0.5cm}
\begin{widetext}
\appendix*
\section{Witness Inequalities}\label{appendix}

\textbf{Multi-cavity witness: Scalar fields}\\

The complete witness inequality that is employed in Eq.~(\ref{eq:BosonsA GME witness})
in Sec.~\ref{subsec:scenario A bosons} is given by
\begin{footnotesize}
\begin{align}
    2\,\Bigl(|\left\langle\right.1_{C}\left.\right|
     \left\langle\right.\tilde{2}_{k\pr}\left.\right|
     \left\langle\right.\tilde{2}_{k}\left.\right|
     \left\langle\right.1_{A}\left.\right|\,
     \tilde{\rho}^{\,\pm}_{Akk\pr C}\,
     \left|\right.\tilde{0}\left.\right\rangle|
     &\,-\,\sqrt{\bra{\!1_{C}\!}
        \tilde{\rho}^{\,\pm}_{Akk\pr C}
        \ket{\!1_{C}\!}
        \bra{\!\tilde{2}_{k\pr}\!}\bra{\!\tilde{2}_{k}\!}\bra{\!1_{A}\!}
        \tilde{\rho}^{\,\pm}_{Akk\pr C}
        \ket{\!1_{A}\!}\ket{\!\tilde{2}_{k}\!}\ket{\!\tilde{2}_{k\pr}\!}}
    \label{eq:BosonsA GME witness complete}\\[1.0mm]
    -\,\sqrt{\bra{\!1_{A}\!}
        \tilde{\rho}^{\,\pm}_{Akk\pr C}
        \ket{\!1_{A}\!}
        \bra{\!1_{C}\!}\bra{\!\tilde{2}_{k\pr}\!}\bra{\!\tilde{2}_{k}\!}
        \tilde{\rho}^{\,\pm}_{Akk\pr C}
        \ket{\!\tilde{2}_{k}\!}\ket{\!\tilde{2}_{k\pr}\!}\ket{\!1_{C}\!}}
    &\,-\,\sqrt{\bra{\!\tilde{2}_{k\pr}\!}
        \tilde{\rho}^{\,\pm}_{Akk\pr C}
        \ket{\!\tilde{2}_{k\pr}\!}
        \bra{\!1_{C}\!}\bra{\!\tilde{2}_{k}\!}\bra{\!1_{A}\!}
        \tilde{\rho}^{\,\pm}_{Akk\pr C}
        \ket{\!1_{A}\!}\ket{\!\tilde{2}_{k}\!}\ket{\!1_{C}\!}}\nonumber\\[1.0mm]
    -\,\sqrt{\bra{\!\tilde{2}_{k}\!}
        \tilde{\rho}^{\,\pm}_{Akk\pr C}
        \ket{\!\tilde{2}_{k}\!}
        \bra{\!1_{C}\!}\bra{\!\tilde{2}_{k\pr}\!}\bra{\!1_{A}\!}
        \tilde{\rho}^{\,\pm}_{Akk\pr C}
        \ket{\!1_{A}\!}\ket{\!\tilde{2}_{k\pr}\!}\ket{\!1_{C}\!}}
    &\,-\,\sqrt{\bra{\!1_{C}\!}\bra{\!\tilde{2}_{k\pr}\!}
        \tilde{\rho}^{\,\pm}_{Akk\pr C}
        \ket{\!\tilde{2}_{k\pr}\!}\ket{\!1_{C}\!}
        \bra{\!\tilde{2}_{k}\!}\bra{\!1_{A}\!}
        \tilde{\rho}^{\,\pm}_{Akk\pr C}
        \ket{\!1_{A}\!}\ket{\!\tilde{2}_{k}\!}}\nonumber\\[1.0mm]
    -\,\sqrt{\bra{\!1_{C}\!}\bra{\!\tilde{2}_{k}\!}
        \tilde{\rho}^{\,\pm}_{Akk\pr C}
        \ket{\!\tilde{2}_{k}\!}\ket{\!1_{C}\!}
        \bra{\!\tilde{2}_{k\pr}\!}\bra{\!1_{A}\!}
        \tilde{\rho}^{\,\pm}_{Akk\pr C}
        \ket{\!1_{A}\!}\ket{\!\tilde{2}_{k\pr}\!}}
    &\,-\,\sqrt{\bra{\!1_{C}\!}\bra{\!1_{A}\!}
        \tilde{\rho}^{\,\pm}_{Akk\pr C}
        \ket{\!1_{A}\!}\ket{\!1_{C}\!}
        \bra{\!\tilde{2}_{k\pr}\!}\bra{\!\tilde{2}_{k}\!}
        \tilde{\rho}^{\,\pm}_{Akk\pr C}
        \ket{\!\tilde{2}_{k}\!}\ket{\!\tilde{2}_{k\pr}\!}}\,\Bigr)
        \,\leq\,0\,.\nonumber
\end{align}
\end{footnotesize}
\vspace*{0.1cm}

\textbf{Multi-cavity witness: Dirac fields}\\

The complete form of the witness of Eq.~(\ref{eq:FermionsA GME witness}) from Sec.~\ref{subsec:scenario A fermions} is given by
\begin{footnotesize}
\begin{align}
    \left|^{\raisebox{0.7pt}{\scriptsize{$-$}}\!\!\!}\fbra{\!\tilde{1}_{\kappa\pr}\!}
     ^{\,+\!\!\!}\fbra{\!\tilde{1}_{\kappa}\!}
     \tilde{\varrho}^{\,\pm}_{A\kappa\kappa\pr C}
     \fket{\!\tilde{1}_{\kappa\pr}\!}^{\!\!-}
     \fket{\!1_{C}\!}^{\!\!+}\right|
     &-\,
     \sqrt{^{\raisebox{1.2pt}{\scriptsize{$+$}}\!\!\!}\fbra{\!1_{C}\!}
        \tilde{\varrho}^{\,\pm}_{A\kappa\kappa\pr C}
        \fket{\!1_{C}\!}^{\!\!+\,+\!\!\!}
        \fbra{\!1_{C}\!}\!
        ^{\raisebox{1.5pt}{\scriptsize{$-$}}\!\!\!}\fbra{\!\tilde{1}_{\kappa\pr}\!}\!
        ^{\raisebox{1.5pt}{\scriptsize{$-$}}\!\!\!}\fbra{\!1_{A}\!}
        \tilde{\varrho}^{\,\pm}_{A\kappa\kappa\pr C}
        \fket{\!1_{A}\!}^{\!\!-}\!
        \fket{\!\tilde{1}_{\kappa\pr}\!}^{\!\!-}\!
        \fket{\!1_{C}\!}^{\!\!+}}
    \label{eq:FermionsA GME witness complete}\\[1mm]
    +\,\left|^{\raisebox{0.7pt}{\scriptsize{$+$}}\!\!\!}\fbra{\!1_{C}\!}
      ^{-\!\!\!}\fbra{\!1_{A}\!}
     \tilde{\varrho}^{\,\pm}_{A\kappa\kappa\pr C}
     \fket{\!\tilde{1}_{\kappa\pr}\!}^{\!\!-}
     \fket{\!1_{C}\!}^{\!\!+}\right|
    &-\,\sqrt{^{\raisebox{1.2pt}{\scriptsize{$-$}}\!\!\!}\fbra{\!\tilde{1}_{\kappa\pr}\!}
        \tilde{\varrho}^{\,\pm}_{A\kappa\kappa\pr C}
        \fket{\!\tilde{1}_{\kappa\pr}\!}^{\!\!\raisebox{-0.4pt}{\scriptsize{$-$}}\,\raisebox{-0.3pt}{\scriptsize{$+$}}\!\!\!}
        \fbra{\!1_{C}\!}\!
        ^{\raisebox{1.0pt}{\scriptsize{$-$}}\!\!\!}\fbra{\!\tilde{1}_{\kappa\pr}\!}\!
        ^{\raisebox{1.2pt}{\scriptsize{$+$}}\!\!\!}\fbra{\!\tilde{1}_{\kappa}\!}
        \tilde{\varrho}^{\,\pm}_{A\kappa\kappa\pr C}
        \fket{\!\tilde{1}_{\kappa}\!}^{\raisebox{0.5pt}{\!\!\scriptsize{$+$}}}\!
        \fket{\!\tilde{1}_{\kappa\pr}\!}^{\raisebox{0.5pt}{\!\!\scriptsize{$-$}}}\!
        \fket{\!1_{C}\!}^{\raisebox{0.4pt}{\!\!\scriptsize{$+$}}}
    }\nonumber
\end{align}
\begin{equation*}
    \ -\,\sqrt{^{\raisebox{1.2pt}{\scriptsize{$+$}}\!\!\!}\fbra{\!1_{C}\!}\!
     ^{\raisebox{1.5pt}{\scriptsize{$-$}}\!\!\!}\fbra{\!\tilde{1}_{\kappa\pr}\!}
     \tilde{\varrho}^{\,\pm}_{A\kappa\kappa\pr C}
     \fket{\!\tilde{1}_{\kappa\pr}\!}^{\!\!-}\!
     \fket{\!1_{C}\!}^{\!\!+}
     \Bigl(\,
        ^{\raisebox{1.0pt}{\scriptsize{$-$}}\!\!\!}\fbra{\!\tilde{1}_{\kappa\pr}\!}\!
        ^{\raisebox{1.2pt}{\scriptsize{$+$}}\!\!\!}\fbra{\!\tilde{1}_{\kappa}\!}
        \tilde{\varrho}^{\,\pm}_{A\kappa\kappa\pr C}
        \fket{\!\tilde{1}_{\kappa}\!}^{\raisebox{0.5pt}{\!\!\scriptsize{$+$}}}\!
        \fket{\!\tilde{1}_{\kappa\pr}\!}^{\raisebox{0.5pt}{\!\!\scriptsize{$-$}}}
     \,+\,
     ^{\raisebox{1.2pt}{\scriptsize{$+$}}\!\!\!}\fbra{\!1_{C}\!}\!
      ^{\raisebox{1.5pt}{\scriptsize{$-$}}\!\!\!}\fbra{\!1_{A}\!}
     \tilde{\varrho}^{\,\pm}_{A\kappa\kappa\pr C}
     \fket{\!_{A}\!}^{\!\!-}\!
     \fket{\!1_{C}\!}^{\!\!+}
     \Bigr)}\,\leq   0\,.
\end{equation*}
\end{footnotesize}

\vspace*{-0.5cm}
\textbf{Single-cavity witness: Scalar fields}\\

In Eq.(\ref{eq:BosonsB GME witness}) in Sec.~\ref{subsec:scenario B bosons} we use the witness inequality
\vspace*{-0.2cm}
\begin{align}
    2\,\Bigl(\,\left|\bra{\tilde{0}}
    \tilde{\rho}_{k\,k\pr k\prpr}
    \ket{\!\tilde{1}_{k}\!}\ket{\!\tilde{2}_{k\pr}\!}\ket{\!\tilde{1}_{k\prpr}\!}\right|
    &\,-\,\sqrt{\bra{\!\tilde{1}_{k}\!}
        \tilde{\rho}_{k\,k\pr k\prpr}
        \ket{\!\tilde{1}_{k}\!}
        \bra{\!\tilde{1}_{k\prpr}\!}\bra{\!\tilde{2}_{k\pr}\!}
        \tilde{\rho}_{k\,k\pr k\prpr}
        \ket{\!\tilde{2}_{k\pr}\!}\ket{\!\tilde{1}_{k\prpr}\!}}
    \label{eq:BosonsB GME witness complete}\\
    -\,\sqrt{\bra{\!\tilde{1}_{k\prpr}\!}
        \tilde{\rho}_{k\,k\pr k\prpr}
        \ket{\!\tilde{1}_{k\prpr}\!}
        \bra{\!\tilde{2}_{k\pr}\!}\bra{\!\tilde{1}_{k}\!}
        \tilde{\rho}_{k\,k\pr k\prpr}
        \ket{\!\tilde{1}_{k}\!}\ket{\!\tilde{2}_{k\pr}\!}}
    &\,-\,\sqrt{\bra{\!\tilde{2}_{k\pr}\!}
        \tilde{\rho}_{k\,k\pr k\prpr}
        \ket{\!\tilde{2}_{k\pr}\!}
        \bra{\!\tilde{1}_{k\prpr}\!}\bra{\!\tilde{1}_{k}\!}
        \tilde{\rho}_{k\,k\pr k\prpr}
        \ket{\!\tilde{1}_{k}\!}\ket{\!\tilde{1}_{k\prpr}\!}}
    \,\Bigr)\,\leq0\,,\nonumber
\end{align}
where the first term is quadratic in $h$ while all other terms are of higher order.\\
\vspace*{0.1cm}

\textbf{Single-cavity witness: Dirac fields}\\

Finally, for the entanglement between three fermionic modes in a single cavity, Sec.~\ref{subsec:scenario B fermions},
the complete form of the witness used in Eq.~(\ref{eq:FermionsB witness}) is
\vspace*{-0.3cm}
\begin{equation}
   \left|\fbra{\tilde{0}}
    \tilde{\varrho}_{\kappa\,\kappa\pr\kappa\prpr}
    \fket{\!\tilde{1}_{\kappa}\!}^{\!\!+}\fket{\!\tilde{1}_{\kappa\prpr}\!}^{\!\!-}\right|\,+\,
    \left|\fbra{\tilde{0}}
    \tilde{\varrho}_{\kappa\,\kappa\pr\kappa\prpr}
    \fket{\!\tilde{1}_{\kappa\pr}\!}^{\!\!+}\fket{\!\tilde{1}_{\kappa\prpr}\!}^{\!\!-}\right|
    \label{eq:FermionsB witness complete}
\end{equation}
\vspace*{-7.5mm}
\begin{equation*}
   -\,\sqrt{
    \fbra{\tilde{0}}
    \tilde{\varrho}_{\kappa\,\kappa\pr\kappa\prpr}
    \fket{\tilde{0}}
    \Bigl(
    ^{\raisebox{-3.7pt}{\scriptsize{$-$}}\!\!}\fbra{\!1_{\kappa\prpr}\!}^{+\!\!\!}\fbra{\!1_{\kappa}\!}
    \tilde{\varrho}_{\kappa\,\kappa\pr\kappa\prpr}
    \fket{\!\tilde{1}_{\kappa}\!}^{\!\!+}\fket{\!\tilde{1}_{\kappa\prpr}\!}^{\!\!-}
    \,+\,
    ^{\raisebox{2.5pt}{\scriptsize{$-$}}\!\!\!}\fbra{\!\tilde{1}_{\kappa\prpr}\!}^{+\!\!\!}\fbra{\!\tilde{1}_{\kappa\pr}\!}
    \tilde{\varrho}_{\kappa\,\kappa\pr\kappa\prpr}
    \fket{\!\tilde{1}_{\kappa\pr}\!}^{\!\!+}\fket{\!\tilde{1}_{\kappa\prpr}\!}^{\!\!-}
    \Bigr)}
    \nonumber
\end{equation*}
\vspace*{-7.5mm}
\begin{equation*}
    -\,\sqrt{^{\raisebox{2.5pt}{\scriptsize{$-$}}\!\!\!}\fbra{\!\tilde{1}_{\kappa\prpr}\!}
    \tilde{\varrho}_{\kappa\,\kappa\pr\kappa\prpr}
    \fket{\!\tilde{1}_{\kappa\prpr}\!}^{\!\!-\,+\!\!}
    \fbra{\!\tilde{1}_{\kappa\pr}\!}
    \tilde{\varrho}_{\kappa\,\kappa\pr\kappa\prpr}
    \fket{\!\tilde{1}_{\kappa\pr}\!}^{\!\!+}
    }\,
    -\,\sqrt{^{\raisebox{2.5pt}{\scriptsize{$-$}}\!\!\!}\fbra{\!\tilde{1}_{\kappa\prpr}\!}
    \tilde{\varrho}_{\kappa\,\kappa\pr\kappa\prpr}
    \fket{\!\tilde{1}_{\kappa\prpr}\!}^{\!\!-\,+\!\!}
    \fbra{\!\tilde{1}_{\kappa}\!}
    \tilde{\varrho}_{\kappa\,\kappa\pr\kappa\prpr}
    \fket{\!\tilde{1}_{\kappa}\!}^{\!\!+}}\,\leq 0\,.
    \nonumber
\end{equation*}
In contrast to Eqs.~({\ref{eq:BosonsA GME witness complete}})-(\ref{eq:BosonsB GME witness complete}) this witness is
constructed using the techniques from Ref.~\cite{HuberMintertGabrielHiesmayr2010} but it presents a lower bound
for the same measures of genuine multipartite entanglement as
Eqs.~({\ref{eq:BosonsA GME witness complete}})-(\ref{eq:BosonsB GME witness complete}).
\end{widetext}

\end{document}